\documentclass[12pt]{article}
\makeatletter
\def\thesectionf{\arabic{section}}
\def\thesection{\Roman{section}}
\def\thesubsection{\Roman{section}.\arabic{subsection}}
\@addtoreset{equation}{section}
\def\theequation{\thesectionf.\arabic{equation}}
\def\appendix{\par
\setcounter{section}{0}
\setcounter{subsection}{0}
\def\thesection{\Alph{section}}
\def\thesubsection{\Alph{section}.\arabic{subsection}}
\def\theequation{\thesection.\arabic{equation}}}
\makeatother
\makeatletter
\def\abstract#1{\long\def\@abstract{#1}}%
\def\@abstract{}%
\let\@oldmaketitle=\@maketitle%
\def\@maketitle{%
\@oldmaketitle%
\begin{center}\large\bf Abstract\end{center}%
\begin{quotation}\@abstract\end{quotation}%
\vskip 1.5em}%
\makeatother
\makeatletter
\def\eqnarray{%
\stepcounter{equation}%
\let\@currentlabel=\theequation
\global\@eqnswtrue
\global\@eqcnt\z@
\tabskip\@centering
\let\\=\@eqncr
$$\halign to \displaywidth\bgroup\@eqnsel\hskip\@centering
$\displaystyle\tabskip\z@{##}$&\global\@eqcnt\@ne
\hfil$\displaystyle{{}##{}}$\hfil
&\global\@eqcnt\tw@$\displaystyle\tabskip\z@{##}$\hfil
\tabskip\@centering&\llap{##}\tabskip\z@\cr}
\makeatother

\newcommand{\measure}[2]{{d\mu\!\left({#1},{#2}\right)}}
\newcommand{\bra}[1]{{\langle{#1}\vert}}
\newcommand{\ket}[1]{{\vert{#1}\rangle}}
\newcommand{\braket}[2]{{\langle{#1}\vert{#2}\rangle}}
\newcommand{\kansu}[2]{{{#1}\!\left({#2}\right)}}
\newcommand{\kakko}[1]{{\left({#1}\right)}}

\newcommand{\wa}[2]{\sum^{#1}_{#2}}
\newcommand{\ffz}{\mbox{\boldmath$z$}}
\newcommand{\flambda}{\mbox{\boldmath$\lambda$}}
\newcommand{\ckk}[1]{\kansu{{}_0F_1}{#1}}
\newcommand{\cph}{{{\mathbf C}P(H)}}
\newcommand{\futoc}{{\mathbf C}}
\begin{document}

\title{\sl Extension of the Barut-Girardello Coherent State\\
        and Path Integral {\rm II}}
\author{
  Kazuyuki FUJII\thanks{e-mail address : fujii@yokohama-cu.ac.jp}
  and
  Kunio FUNAHASHI\thanks{e-mail address : funahasi@yokohama-cu.ac.jp }
  \\
  Department of Mathematics, Yokohama City University,\\
  Yokohama 236, Japan}
\date{August, 1997}

\abstract{
We have constructed the coherent state of $U(N,1)$ , which is an
extension of the Barut-Girardello (BG) coherent state of $SU(1,1)$,
in our previous paper.
However there is a restriction that the eigenvalue of the Casimir
operator is natural number.
In this paper we construct the coherent state in the analytic
representation to overcome this restriction.
Next we show that the measure of the BG coherent state is not the
symplectic induced measure.
}
\maketitle\thispagestyle{empty}
\newpage

\section{Introduction}
\label{sec:jo}

Coherent state of the harmonic oscillator is defined as the eigenstate
of the annihilation operator and has been utilized for revealation of many
physical properties.
Concurrently its definition has been extended~\cite{KLAUDER}.

As a straightforward extension of the definition, there exists the
Barut-Girardello (BG) coherent state~\cite{BG}, which is defined as the
eigenstate of the lowering operator in $SU(1,1)$.
The remarkable property is that the range of the eigenvalue of the
Casimir operator is $K>0$, in spite of the representation of $SU(1,1)$
being defined for $K\ge1/2$.
(From this fact, the BG coherent state may be the coherent state of some
covering group of $SU(1,1)$.)
According to some groups there are further extensions of the BG coherent
state~\cite{DQ,RON:BGCS,TRIFONOV}.
In our previous work~\cite{RON:BGCS}, we have constructed the extended
BG coherent state based on some representation of $U(N,1)$ and
its measure.
However its eigenvalue of the Casimir, $K$, is restricted to natural number
because the Schwinger boson method~\cite{SCHWINGER}  is used in the
construction.
Thus in this paper we construct the coherent state in the analytic
representation to overcome this restriction.

Although the BG coherent state is a straghtforward extension of that of
the harmonic oscillator, the measure is given by the integral
formula~\cite{BATEMAN}.
While, ordinary measures such as the harmonic oscillator or the Perelomov
coherent state~\cite{PERELOMOV} are ones induced from the canonical
symplectic 2-form on the infinite dimensional complex projective space
(hereafter abbreviated as the symplectic induced measure).
Thus we investigate whether the measure of the BG coherent state
(hereafter abbreviated as the BG measure) is
the symplectic induced measure or not.

The contents of this paper are as follows.
In Sec \ref{sec:holo} we construct the extended BG coherent state
in the analytic representation.
In Sec \ref{sec:measure} we show that the BG measure is not
the symplectic induced measure.
The last section is devoted to the discussion.

\section{Analytic Representation of the BG Coherent State}
\label{sec:holo}

We review the BG coherent state in \ref{sec:bgmoto} and construct the
extended coherent state in the analytic representation
in \ref{sec:bgkakuchou}.

\subsection{The BG coherent state}
\label{sec:bgmoto}

$su(1,1)$ algebra satisfies
\begin{equation}
  \label{holo:daisu}
  [K_3,K_\pm]=\pm K_\pm\ ,\
  [K_-,K_+]=2K_3\ ,\
  \kakko{K_\pm=\pm\kakko{K_1\pm iK_2}}\ ,
\end{equation}
and the representation is
\begin{equation}
    \{
  \ket{K,m}\vert m=0,1,2,\cdots
  \}\ ,\
  K\ge{1\over2}\ ,\
  \textrm{($2K$ is an eigenvalue of the Casimir operator)}\ .
\end{equation}
They satisfy
\begin{eqnarray}
  \label{holo:suii}
  K_3\ket{K,m}
  &=&
  (K+m)\ket{K,m}\ ,
  \nonumber\\
  K_+\ket{K,m}
  &=&
  \sqrt{(m+1)(2K+m)}\ket{K,m+1}\ ,
  \nonumber\\
  K_-\ket{K,m}
  &=&
  \sqrt{m(2K+m-1)}\ket{K,m-1}\ .
\end{eqnarray}

The BG coherent state is defined as the eigenstate of the lowering operator:
\begin{equation}
  \label{holo:bgteigi}
  K_-\ket{z}=z\ket{z}\ .
\end{equation}
The explicit form of (\ref{holo:bgteigi}) is
\begin{equation}
  \label{cs:bgcs}
  \ket{z}=\wa{\infty}{n=0}{z^n\over\sqrt{n!\kakko{2K}_n}}\ket{K,n}\ ,\
z\in{\bf C}\ .
\end{equation}
The inner product is
\begin{equation}
  \braket{z}{z^\prime}
  =
  \kansu{\Gamma}{2K}\kakko{z^*z^\prime}^{-K+{1\over2}}
  \kansu{I_{2K-1}}{2\sqrt{z^*z^\prime}}
  =
  \ckk{2K;z^*z^\prime}\ ,
\end{equation}
where $I_\nu(z)$ is a modified Bessel function of the first kind
defined in (\ref{measure:besselteigi}) and $\ckk{\nu;z}$ is defined in
(\ref{holo:ckkteigi}).
The resolution of unity is
\begin{eqnarray}
  \label{holo:bgcsrou}
  \int\kansu{d\mu}{z,z^*}\ket{z}\bra{z}=1_K\ ,
  \nonumber\\
  \kansu{d\mu}{z,z^*}
  \equiv
  {2\kansu{K_{2K-1}}{2\vert z\vert}\over\pi\kansu{\Gamma}{2K}}
  \vert z\vert^{2K-1}dz^*dz\ ,
\end{eqnarray}
where $1_K$ is the identity operator in the representation space.
It is remarkable that (\ref{holo:bgcsrou}) holds for $K>0$.

So far we have expressed the BG coherent state by means of the Dirac
notation.
Alternatively it is possible to express in the analytic representation.
We adopt the bases as
\begin{equation}
  \label{holo:bgkitei}
  u_n
  \equiv
  \sqrt{1\over n!\kakko{2K}_n}z^n\
  \kakko{=\ket{K,n}}\ ,
\end{equation}
where
\begin{equation}
    (a)_n\equiv a\cdot(a+1)\cdots(a+n-1)\ ,
\end{equation}
and the operators as
\begin{equation}
  K_+ = z\ ,\
  K_- = z{d^2\over dz^2}+2K{d\over dz}\ ,\
  K_3 = z{d\over dz}+K\ .
\end{equation}
Of course, they satisfy (\ref{holo:daisu}) and (\ref{holo:suii}).
Eq (\ref{holo:bgkitei}) satisfies the completeness
\begin{equation}
  \label{holo:kanzen}
  \wa{\infty}{n=0}\kansu{u_n}{z^\prime}\kansu{u_n^*}{z}
  =
  \ckk{2K;z^\prime z^*}\ ,
\end{equation}
where $\ckk{\nu;z}$ is the hypergeometric function:
\begin{equation}
  \label{holo:ckkteigi}
  \ckk{\nu;z}
  \equiv
  \wa{\infty}{n=0}{1\over\kakko{\nu}_n}{z^n\over n!}\ .
\end{equation}
The inner product is defined by
\begin{equation}
  \label{holo:bgnaiseki}
  \kakko{A,B}
  \equiv
  \int\measure{z}{z^*}\kansu{A^*}{z}\kansu{B}{z}\ ,\
  \textrm{($A$, $B$: analytic functions)}\ ,
\end{equation}
where
\begin{equation}
  \measure{z}{z^*}
  \equiv
  {2\kansu{K_{2K-1}}{2\vert z\vert}\over\pi\kansu{\Gamma}{2K}}
  \vert z\vert^{2K-1}dz^*dz\ .
\end{equation}
Then the BG coherent state is written by
\begin{eqnarray}
  \label{holo:bgcs}
  &&
  \kansu{\varphi}{\lambda}
  \equiv
  \ckk{2K;\lambda z}
  =
  \wa{\infty}{n=0}{1\over n!\kakko{2K}_n}
  \kakko{\lambda z}^n\ ,
  \nonumber\\
  &&
  K_-\kansu{\varphi}{\lambda}=\lambda\kansu{\varphi}{\lambda}\ ,
\end{eqnarray}
which satisfies, of course, all of the properties of the
BG coherent state.

\subsection{Extension of the BG coherent state}
\label{sec:bgkakuchou}

$u(N,1)$ algebra is defined by
\begin{eqnarray}
  \label{ncp:daisuteigi}
  &&
  [E_{\alpha\beta},E_{\gamma\delta}]
  =
  \eta_{\beta\gamma}E_{\alpha\delta}
  -\eta_{\delta\alpha}E_{\gamma\beta}\ ,
  \nonumber\\
  &&
  \eta_{\alpha\beta}
  =
  \kansu{\rm diag}{1,\cdots,1,-1}\ ,\
  \kakko{\alpha,\beta,\gamma,\delta = 1,\cdots,N+1}\ ,
\end{eqnarray}
with a subsidiary condition
\begin{equation}
  -\wa{N}{\alpha=1}E_{\alpha\alpha}+E_{N+1,N+1}=K\ ,\
  \kakko{K=1,2,\cdots}\ .
\end{equation}
Now we briefly review the extension by means of the Schwinger boson
method~\cite{SCHWINGER}.
We identify these generators with creation and annihilation operators
of harmonic oscillators:
\begin{equation}
  \matrix{
    &E_{\alpha\beta}
    =a_\alpha^\dagger a_\beta\ ,&
    E_{\alpha,N+1}
    =a_\alpha^\dagger a_{N+1}^\dagger\ ,
    \cr
    &E_{N+1,\alpha}
    =a_{N+1}a_\alpha\ ,&
    E_{N+1,N+1}
    =a_{N+1}^\dagger a_{N+1}+1\ ,
    \cr
    }
\end{equation}
where $a$, $a^\dagger$ satisfy
\begin{equation}
  [a_\alpha,a_\beta^\dagger]=1\ ,\
  [a_\alpha,a_\beta]=[a_\alpha^\dagger,a_\beta^\dagger]=0\ ,\
  \kakko{\alpha,\beta=1,2,\cdots,N+1}\ .
\end{equation}
The Fock space is
\begin{eqnarray}
  &&
  \left\{
    \ket{n_1,\cdots,n_{N+1}}\vert n_1,n_2,\cdots,n_{N+1}=0,1,2,\cdots
  \right\}\ ,
  \nonumber\\
  \ket{n_1,\cdots,n_{N+1}}
  &\equiv&
  {1\over\sqrt{n_1!\cdots n_{N+1}!}}
  \kakko{a_1^\dagger}^{n_1}\cdots\kakko{a_{N+1}^\dagger}^{n_{N+1}}
  \ket{0,0,\cdots,0}\ ,
  \nonumber\\
  &&
  a_\alpha\ket{0,0,\cdots,0}=0\ .
\end{eqnarray}
On the representation space it is
\begin{equation}
  1_K
  \equiv
  \wa{\infty}{\left\{ n\right\}=0}
  \ket{n_1,\cdots,n_N,K-1+\sum^N_{\alpha=1}n_\alpha}
  \bra{n_1,\cdots,n_N,K-1+\sum^N_{\alpha=1}n_\alpha}\ ,
\end{equation}
where an abbreviation
\begin{equation}
  \sum^\infty_{\left\{ n\right\}=0}
  \equiv
  \wa{\infty}{n_1=0}
  \wa{\infty}{n_2=0}
  \cdots
  \wa{\infty}{n_N=0}\ ,
\end{equation}
has been used.

Then the coherent state is defined by
\begin{equation}
  E_{N+1,\alpha}\ket{\ffz}
  =
  z_\alpha\ket{\ffz}\ ,\ \kakko{\alpha=1,\cdots,N}\ ,
\end{equation}
and the explicit form is
\begin{equation}
  \label{holo:kata}
  \ket{\ffz}
  =
  \wa{\infty}{\left\{ n\right\}=0}
  \sqrt{\kansu{\Gamma}{K}\over
    n_1!\ldots n_N!\kansu{\Gamma}{K+\wa{N}{\beta=1}n_\beta}}
  z_1^{n_1}\ldots z_N^{n_N}
  \ket{n_1,\ldots,n_N,K-1+\wa{N}{\alpha=1}n_\alpha}\ .
\end{equation}
Their inner product is
\begin{equation}
  \label{bgcs:naiseki}
  \braket{\ffz}{\ffz^\prime}
  =
  \ckk{K;\ffz^\dagger\ffz^\prime}\ ,\
  \kakko{\ffz^\dagger\ffz^\prime\equiv
    z_1^*z_1^\prime+\cdots+z_N^*z_N^\prime}\ ,
\end{equation}
where $\ckk{\nu;z}$ is defined in (\ref{holo:ckkteigi}),
and the resolution of unity is
\begin{equation}
  \int\measure{\ffz}{\ffz^\dagger}\ket{\ffz}\bra{\ffz}=1_K\ ,
\end{equation}
where
\begin{eqnarray}
  \kansu{d\mu}{\ffz,\ffz^\dagger}
  &\equiv&
  {2\|\ffz\|^{K-N}\kansu{K_{K-N}}{2\|\ffz\|}\over\pi^N\kansu{\Gamma}{K}}
  [d\ffz^\dagger d\ffz]\ ,
  \nonumber\\
   \|\ffz\|
   &\equiv&
   \sqrt{\ffz^\dagger\ffz}\ ,\
   [d\ffz^\dagger d\ffz]
   \equiv
   \prod^N_{\alpha=1}
   \kansu{d}{{\rm Re}z_\alpha}\kansu{d}{{\rm Im}z_\alpha}\ .
\end{eqnarray}

In this expression the representation of the harmonic oscillator restricts
$K$ to natural number.
Thus we write the coherent state by means of the analytic representation
to overcome this restriction.

When we adopt the bases as
\begin{equation}
  \label{holo:gkitei}
  u_{n_1,\cdots,n_N}
  \equiv
  {1\over\sqrt{n_1!\cdots n_N!\kakko{K}_{\wa{N}{\alpha=1}n_\alpha}}}
  z_1^{n_1}\cdots z_N^{n_N}\ ,
\end{equation}
the operators are written as
\begin{eqnarray}
  E_{\alpha\beta} = z_\alpha{\partial\over\partial z_\beta}\ ,\
  E_{N+1,N+1} = \wa{N}{\alpha=1}z_\alpha{\partial\over\partial z_\alpha}+K\ ,
  \nonumber\\
  E_{\alpha,N+1} = z_\alpha\ ,\
  E_{N+1,\alpha} = \wa{N}{\beta=1}z_\beta{\partial^2\over\partial z_\beta
    \partial z_\alpha}+K{\partial\over\partial z_\alpha}\ ,
  \nonumber\\
  \kakko{\alpha=1,\cdots,N}\ .
\end{eqnarray}
Eq (\ref{holo:gkitei}) satisfies the completeness
\begin{equation}
  \wa{\infty}{\{ n\}=0}
  \kansu{u_{n_1,\cdots,n_N}}{z^\prime}
  \kansu{u_{n_1,\cdots,n_N}^*}{z}
  =
  \ckk{K;\ffz^\prime\cdot\ffz^*}\ ,
\end{equation}
where the dot is defined as
\begin{equation}
  \ffz^\prime\cdot\ffz^*
  \equiv
  z_1^\prime z_1^*+\cdots+z_N^\prime z_N^*\ .
\end{equation}
The inner product is defined by
\begin{equation}
  \kakko{A,B}
  \equiv
  \int\measure{\ffz}{\ffz^\dagger}
  \kansu{A^*}{\ffz}\kansu{B}{\ffz}\ ,\
  \textrm{($A$, $B$: analytic functions)}\ ,
\end{equation}
where
\begin{equation}
  \measure{\ffz}{\ffz^\dagger}
  \equiv
  {2\|\ffz\|^{K-N}\kansu{K_{K-N}}{2\|\ffz\|}\over
    \pi^N\kansu{\Gamma}{K}}
  [d\ffz^\dagger d\ffz]\ .
\end{equation}
Then the coherent state is defined by
\begin{equation}
  E_{N+1,\alpha}\kansu{\varphi}{\flambda}
  =
  \lambda_\alpha\kansu{\varphi}{\flambda}\ ,
\end{equation}
and whose explicit form is obtained from (\ref{holo:kata}) as
\begin{eqnarray}
  \kansu{\varphi}{\flambda}
  &=&
  \wa{\infty}{\{ n\}=0}
  {1\over\sqrt{n_1!\cdots n_N!\kakko{K}_{\wa{N}{\alpha=1}n_\alpha}}}
  \lambda_1^{n_1}\cdots\lambda_N^{n_N}u_n
  \nonumber\\
  &=&
  \wa{\infty}{\{ n\}=0}
  {1\over{n_1!\cdots n_N!\kakko{K}_{\wa{N}{\alpha=1}n_\alpha}}}
  \kakko{\lambda_1z_1}^{n_1}\cdots\kakko{\lambda_Nz_N}^{n_N}
  \nonumber\\
  &=&
  \wa{\infty}{m=0}{1\over m!\kakko{K}_m}
  \wa{}{n_1+\cdots+n_N=m}
  {m!\over n_1!\cdots n_N!}
  \kakko{\lambda_1z_1}^{n_1}\cdots\kakko{\lambda_Nz_N}^{n_N}
  \nonumber\\
  &=&
  \ckk{K;\flambda\cdot\ffz}\ .
\end{eqnarray}
This is the analytic representation of the extended BG coherent
state, which no longer restricts $K$ to natural number.

\section{The Measure of the BG Coherent State}
\label{sec:measure}

In this section, first we show the form of the symplectic induced measure
and then we compare it with the BG measure.

We define the infinite dimensional complex projective space:
\begin{eqnarray}
  \cph
  &\equiv&
  H-\{0\}/\futoc^*\ ,\
  \kakko{\futoc^*\equiv\futoc-\{0\}}\ ,
  \nonumber\\
  H
  &\equiv&
  \kansu{l^2}{\futoc}\ .
\end{eqnarray}
$\cph$ is an infinite dimensional symplectic manifold and its
element is written as
\begin{equation}
  \kansu{P}{X}
  \equiv
  X\kakko{X^\dagger X}^{-1}X^\dagger
  =
  {XX^\dagger\over X^\dagger X}\ ,\
  \kakko{X\in H-\{0\}}\ .
\end{equation}
Then the canonical symplectic 2-form on $\cph$ is given by
\begin{equation}
  \kansu{\omega_\infty}{X}
  \equiv
  \kansu{{\mathrm {Tr}}}{\kansu{P}{X}\kansu{dP}{X}\wedge\kansu{dP}{X}}\ .
\end{equation}

Next we define a map $f: M\to\cph\ (M=D(1,1)\textrm{ or }\futoc)$
such that
\begin{equation}
  \label{fsm:pb}
  \kansu{f}{z}
  =
  {\ket{z}\bra{z}\over\braket{z}{z}}\ ;\
  X=\ket{z}\ .
\end{equation}
By means of the map, we pullback the symplectic 2-form on $\cph$ to $M$:
\begin{equation}
  \label{fsm:omegam}
  \kansu{\omega_M}{z}
  =
  \kansu{{\mathrm {Tr}}}{\kansu{f}{z}\kansu{df}{z}\wedge\kansu{df}{z}}\ ,
\end{equation}
where $d$ is the exterior derivative on $M$.
Putting (\ref{fsm:pb}) into (\ref{fsm:omegam}), we obtain the explicit
form:
\begin{equation}
  \kansu{\omega_M}{z}
  =
  dz^*dz{1\over\braket{z}{z}}{\partial^2\over\partial z^*\partial z}
  \log\braket{z}{z}\ ,
\end{equation}
and this is the symplectic induced measure in 2 dimension.

Usually path integral measures are given  by the symplectic induced measure.
As an example, we consider the Perelomov coherent state in $SU(1,1)$:
\begin{eqnarray}
  \label{measure:perelomov}
  \ket{\xi}
  \equiv
  e^{\xi K_+}\ket{K,0}
  =
  \wa{\infty}{n=0}\sqrt{\kakko{2K}_n\over n!}\xi^n\ket{K,n}\ ,\
  \xi
  &\in&
  \kansu{D}{1,1}\ ,
\end{eqnarray}
where
\begin{equation}
  \kansu{D}{1,1}
  \equiv
  \big\{
  \xi\in{\bf C}\big\vert\vert\xi\vert<1
  \big\}\ .
\end{equation}
The inner product is
\begin{equation}
  \braket{\xi}{\xi^\prime}
  =
  \kakko{1-\xi^*\xi^\prime}^{-2K}\ ,
\end{equation}
and the resolution of unity is
\begin{equation}
  \int\measure{\xi}{\xi^*}\ket{\xi}\bra{\xi}=1_K\ ,
\end{equation}
where the measure is
\begin{equation}
  \label{measure:pcsmeasure}
  \kansu{d\mu}{\xi,\xi^*}
  =
  {2K-1\over\pi}
  {d\xi^*d\xi\over\kakko{1-\vert\xi\vert^2}^{-2K+2}}\ .
\end{equation}
The symplectic induced measure by (\ref{measure:perelomov}) is given by
\begin{eqnarray}
  \label{measure:peremesure}
   \omega
   &\equiv&
   d\xi^*d\xi
   {1\over\braket{\xi}{\xi}}
   {\partial^2\over\partial\xi^*\partial\xi}
   \log\braket{\xi}{\xi}\ ,\
   \kakko{\braket{\xi}{\xi}=\kakko{1-\vert\xi\vert^2}^{-2K}}\ ,
   \nonumber\\
   &=&
   2K{d\xi^*d\xi\over\kakko{1-\vert\xi\vert^2}^{-2K+2}}\ .
\end{eqnarray}
This is quite the same with (\ref{measure:pcsmeasure})
if (\ref{measure:peremesure}) is normalized.

Now, turning to the the BG coherent state, the measure:
\begin{equation}
  \label{measure:bgmeasure}
  \kansu{d\mu}{z,z^*}
  \equiv
  {2\kansu{K_{2K-1}}{2\vert z\vert}\over\pi\kansu{\Gamma}{2K}}
  \vert z\vert^{2K-1}dz^*dz\ ,
\end{equation}
is obtained by the integral formula~\cite{BATEMAN}:
\begin{equation}
  \label{measure:iwanami}
  \int^\infty_0dx2x^{\alpha+\beta}
  \kansu{K_{2\kakko{\alpha-\beta}}}{2x^{1/2}}x^{s-1}
  =
  \kansu{\Gamma}{2\alpha+s}\kansu{\Gamma}{2\beta+s}\ .
\end{equation}
We investigate whether it is the symplectic induced measure or not.
Let us calculate
\begin{equation}
  \label{measure:form}
  \omega
  \equiv
  dz^*dz
  {1\over\braket{z}{z}}{\partial^2\over\partial z^*\partial z}
   \log\braket{z}{z}\ ,\
   \kakko{\braket{z}{z}=\ckk{2K;\vert z\vert^2}}\ .
\end{equation}
Utilizing the polar coordinate
\begin{equation}
  z = re^{i\theta}\ ,
\end{equation}
we write (\ref{measure:form}) as
\begin{equation}
  \label{measure:formhen}
  \omega
  =
  dz^*dz
  {1\over\ckk{2K;r^2}}
  {1\over4}{1\over r}
  {d\over dr}\kakko{r{d\over dr}}
  \log\ckk{2K;r^2}\ .
\end{equation}
By noting
\begin{equation}
  {d\over dt}\ckk{\nu;t}
  =
  {1\over\nu}\ckk{\nu+1;t}\ ,
\end{equation}
eq (\ref{measure:formhen}) becomes
\begin{eqnarray}
  \label{measure:formsaigo}
  \omega
  &=&
  dz^*dz
  {1\over2K}{1\over\left\{\ckk{2K;r^2}\right\}^3}
  \Bigg[\ckk{2K+1;r^2}\ckk{2K;r^2}
  \nonumber\\
  &&
  +{1\over2K+1}r^2\ckk{2K+2;r^2}\ckk{2K;r^2}
  -{1\over2K}r^2\left\{\ckk{2K+1;r^2}\right\}^2\Bigg]\ .
\end{eqnarray}
Eq (\ref{measure:formsaigo}) looks different from the BG measure.
Now we compare the behavior of (\ref{measure:formsaigo}) with that of
the BG measure near the origin.
{}From the definition of the hypergeometric function (\ref{holo:ckkteigi}),
the behavior near the origin in $O(z)$ is
\begin{equation}
  \ckk{\nu;z}\sim1+{z\over\nu}\ .
\end{equation}
Thus the behavior of $\omega$ in $O(r^2)$ is
\begin{equation}
  \label{measure:omegagenten}
  \omega
  \sim
  dz^*dz
  {1\over2K}\left[1-{2K+3\over2K\kakko{2K+1}}r^2\right]\ .
\end{equation}
On the other hand, from the definition of modified Bessel functions:
\begin{eqnarray}
  \label{measure:besselteigi}
  \kansu{K_\nu}{z}
  &=&
  {\pi\over2}{\kansu{I_{-\nu}}{z}-\kansu{I_\nu}{z}\over\sin\nu\pi}\ ,
  \nonumber\\
  \kansu{I_\nu}{z}
  &=&
  \kakko{z\over2}^\nu\wa{\infty}{n=0}
  {\kakko{z/2}^{2n}\over n!\kansu{\Gamma}{\nu+n+1}}\ ,
\end{eqnarray}
the behavior of the BG measure is
\begin{eqnarray}
  \label{measure:bggenten}
  {2\kansu{K_{2K-1}}{2\vert z\vert}\over\pi\kansu{\Gamma}{2K}}
  \vert z\vert^{2K-1}
  &\sim&
  {1\over\kansu{\Gamma}{2K}\sin\kakko{2K-1}\pi}
  \Bigg[\bigg\{{1\over\kansu{\Gamma}{-2K+2}}
  +{r^2\over\kansu{\Gamma}{-2K+3}}+\cdots\bigg\}
  \nonumber\\
  &&
  -r^{K-1/2}\bigg\{{1\over\kansu{\Gamma}{2K}}
  +{r^2\over\kansu{\Gamma}{2K+1}}+\cdots\bigg\}\Bigg]
  \nonumber\\
  &=&
  {1\over\sin\kakko{2K-1}\pi}
  \Bigg[\bigg\{{\sin\kakko{2K-1}\pi\over\kakko{2K-1}\pi}
  -{\sin\kakko{2K-1}\pi\over\kakko{2K-1}\kakko{2K-2}\pi}r^2
  +\cdots\bigg\}
  \nonumber\\
  &&
  -r^{K-1/2}\bigg\{{1\over\kansu{\Gamma}{2K}\kansu{\Gamma}{2K}}
  +{1\over\kansu{\Gamma}{2K}\kansu{\Gamma}{2K+1}}r^2
  +\cdots\bigg\}\Bigg]\ ,
\end{eqnarray}
where we have used the formula
\begin{equation}
  \kansu{\Gamma}{z}\kansu{\Gamma}{1-z}
  =
  {\pi\over\sin\pi z}\ ,
\end{equation}
in the second equality.
For $K>1/2$, (\ref{measure:bggenten}) becomes
\begin{equation}
  \label{measure:bggentenkinji}
  (\ref{measure:bggenten})
  =
  {1\over\kakko{2K-1}\pi}\bigg[1-{1\over2K-2}r^2+\cdots\bigg]\ .
\end{equation}

If $\omega$ is assumed to be another measure of the BG coherent state,
then the resolution of unity with the measure must hold.
On the other hand, $d\mu$ is the measure of the BG coherent
state and satisfies the resolution of unity (\ref{holo:bgcsrou}).
Then by the Lebesgue's theorem, $\omega$ must coincide with $d\mu$.
However, comparing (\ref{measure:omegagenten}) with
(\ref{measure:bggentenkinji}) apart from the total normalization,
the behaviors of them are different.
Thus we conclude that $\omega$ is not a measure of the BG coherent
state, or in other words, that {\em the measure of the BG coherent state
$d\mu$ is not the symplectic induced measure.}

\section{Discussion}
\label{sec:giron}

We have constructed an extended BG coherent state for a representation
of $U(N,1)$ by means of the analytic representation
to overcome the restriction of $K$.

We have shown that the BG measure is not the symplectic induced measure.
This is conclusively different from the Perelomov coherent state,
and may be the reason why calculation of the path integral
becomes difficult in contrast with that of the Perelomov coherent
state~\cite{RON:SUTWO,RON:CPN,GAUSS}.

There are attempts to explain the meaning of the measure of
the coherent state~\cite{PROVOST}.
However all examples in it are the symplectic induced measures.
The BG measure is the first example which is not the symplectic
induced measure.

In spite of our effort, however, the essential meaning of the BG measure
is not still clear.
It is very important to reveal it.

\appendix

\begin{center}
  \Large\bfseries Appendix
\end{center}

\section{Comparison the BG Coherent State to the Symplectic Measure
  in $K=1/4$ and $K=3/4$}

In Sec \ref{sec:measure}, we have shown that the BG measure
is not the symplectic induced measure by comparing them
near the origin.
In this appendix, we compare them in the explicit form for $K=1/4$
and $K=3/4$ to convince that they are really different measures.

\subsection{$K=1/4$ Case}

We put the explicit form of the modified Bessel function:
\begin{equation}
  \kansu{K_{-1/2}}{z}
  =
  \sqrt{\pi\over2z}e^{-z}\ ,
\end{equation}
into the BG measure (\ref{measure:bgmeasure})
to obtain
\begin{equation}
  \label{app:bgkofe}
  \kansu{d\mu_{K=1/4}}{z,z^*}
  =
  dz^*dz{1\over\pi r}e^{-2r}\ ,
\end{equation}
where we have used the polar coordinate ($z=re^{i\theta}$)
except for $dz^*dz$.

On the other hand, the symplectic induced measure (\ref{measure:formsaigo})
for $K=1/4$ is
\begin{eqnarray}
  \label{app:omegakof}
  \omega_{K=1/4}
  &=&
  dz^*dz
  {2\over\Big\{\ckk{{1\over2};r^2}\Big\}^3}
  \Bigg[
  \ckk{{3\over2};r^2}\ckk{{1\over2};r^2}
  +{2\over3}r^2\ckk{{5\over2};r^2}\ckk{{1\over2};r^2}
  \nonumber\\
  &&
  -2r^2\bigg\{\ckk{{3\over2};r^2}\bigg\}^2
  \Bigg]\ .
\end{eqnarray}
Then putting the explicit forms of the hypergeometric functions:
\begin{eqnarray}
  \label{app:fgutai}
  \ckk{{1\over2};r^2}
  &=&
  \cosh2r\ ,
  \nonumber\\
  \ckk{{3\over2};r^2}
  &=&
  {\sinh2r\over2r}\ ,
  \nonumber\\
  \ckk{{5\over2};r^2}
  &=&
  {3\over\kakko{2r}^2}\kakko{\cosh2r-{\sinh2r\over2r}}\ ,
\end{eqnarray}
into (\ref{app:omegakof}), we finally obtain
\begin{equation}
  \label{app:omegakofe}
  \omega_{K=1/4}
  =
  dz^*dz
  {1\over\kakko{\cosh2r}^3}
  \kakko{{\sinh2r\cosh2r\over2r}+1}\ .
\end{equation}
Apparently (\ref{app:omegakofe}) is different from
(\ref{app:bgkofe}).

\subsection{$K=3/4$ Case}

By the explicit form of the modified Bessel function
\begin{equation}
  \kansu{K_{K=1/2}}{z}
  =
  \sqrt{\pi\over2z}e^{-z}\ ,
\end{equation}
we write the BG measure as
\begin{equation}
  \label{app:bgktfe}
  \kansu{d\mu_{K=3/4}}{z,z^*}
  =
  dz^*dz
  {2\over\pi}e^{-2r}\ .
\end{equation}
On the other hand, by the explicit forms of the hypergeometric functions
(\ref{app:fgutai}) and
\begin{equation}
  \ckk{{7\over2};r^2}
  =
  15\kakko{{\sinh2r\over\kakko{2r}^3}
    -3{\cosh2r\over\kakko{2r}^4}
    +3{\sinh2r\over\kakko{2r}^5}}\ ,
\end{equation}
the symplectic induced measure (\ref{measure:formsaigo}) for $K=3/4$ becomes
\begin{eqnarray}
  \label{app:omegaktfe}
  \omega_{3/4}
  &=&
  dz^*dz
  {2\over3}{1\over\Big\{\ckk{{3\over2};r^2}\Big\}^3}
  \Bigg[
  \ckk{{5\over2};r^2}\ckk{{3\over2};r^2}
  +{2\over5}r^2\ckk{{7\over2};r^2}\ckk{{3\over2};r^2}
  \nonumber\\
  &&
  -{2\over3}r^2\bigg\{\ckk{{5\over2};r^2}\bigg\}^2
  \Bigg]
  \nonumber\\
  &=&
  dz^*dz
  {2r\over\kakko{\sinh2r}^3}
  \kakko{{\cosh2r\sinh2r\over2r}-1}\ .
\end{eqnarray}
As well as the $K=1/4$ case, (\ref{app:omegaktfe}) is  different
from (\ref{app:bgktfe}).


\begin{thebibliography}{99}
\bibitem{KLAUDER}
  J.R. Klauder and Bo-S. Skagerstam,
  {\sl Coherent States} (World Scientific, Singapore, 1985).
\bibitem{BG}
  A. O. Barut and L. Girardello,
  Commun. Math. Phys. {\bf 21} (1971) 41.
\bibitem{DQ}
    J. Deene and C. Quesne,
    J. Math. Phys. {\bf 25} (1984) 1638.
\bibitem{RON:BGCS}
  K. Fujii and K. Funahashi,
  to be published in J. Math. Phys.
\bibitem{TRIFONOV}
  D. A. Trifonov,
  preprint INRNE-TH-97/5 (12 June 1997).
\bibitem{SCHWINGER}
  J. Schwinger,
  {\em ON ANGULAR MOMENTUM} in {\em QUANTUM THEORY OF ANGULAR MOMENTUM}
  (Academic press, New York, 1965).
\bibitem{BATEMAN}
  {\em Bateman Project: Vol. I. Integral transformations}, p. 349,
  (MacGrow-Hill, New York, 1954).
\bibitem{PERELOMOV}
  A. Perelomov,
  {\em Generalized Coherent States and Their Applications}
  (Springer-Verlag, Berlin, 1986).
\bibitem{RON:SUTWO}
  K. Funahashi, T. Kashiwa, S. Sakoda and K. Fujii,
  J. Math. Phys. {\bf 36} (1995) 3232.
\bibitem{RON:CPN}
  K. Funahashi, T. Kashiwa, S. Sakoda and K. Fujii,
  J. Math. Phys. {\bf 36} (1995) 4590.
\bibitem{GAUSS}
  K. Fujii, T. Kashiwa and S. Sakoda,
  J. Math. Phys. {\bf 37} (1996) 567.
\bibitem{PROVOST}
  J. P. Provost and G. Vallee,
  Commun. Math. Phys. {\bf 76} (1980) 289.
\end{thebibliography}
\end{document}